\documentclass[a4paper,11pt]{article}
\usepackage{pos}

\def\ubar{\overline{u}}

\def\lbar{\overline{\ell}}

\title{Null test searches for BSM physics with rare charm decays}

\author*[a]{Marcel Golz}

\affiliation[a]{Fakultät für Physik, TU Dortmund\\
 Otto-Hahn-Str. 4, D-44221 Dortmund, Germany}

\emailAdd{marcel.golz@tu-dortmund.de}

\abstract{Semileptonic Flavor Changing Neutral Current decays of charmed hadrons uniquely probe the up-sector of the Standard Model.
Clean null test observables, such as CP-asymmetries, lepton-universality ratios, missing energy modes, lepton flavor violating modes and angular observables can be studied in a variety of charmed meson and baryon modes. An overview of null test opportunities with rare charm decays is presented and similarities and complementarities of different modes are worked out.}

\FullConference{%
  *** 10th International Workshop on Charm Physics (CHARM2020), ***\\
  *** 31 May - 4 June, 2021 ***\\
  *** Mexico City, Mexico - Online ***
}

\begin{document}
\rightline{DO-TH 21/25}

\maketitle

\section{Introduction}
\noindent$\vert \Delta c\vert=\vert\Delta u \vert = 1 $ transitions proceed via 1-loop penguin diagrams with down-type quarks in the loop. A strong Glashow-Iliopoulos-Maiani (GIM) cancellation and Cabbibo-Kobayashi-Maskawa (CKM) suppression then lead to tiny Standard Model (SM) contributions to rare charm decays~\cite{Burdman:2001tf}. Thus, $c\to u\ell^+\ell^-$ induced decays are highly sensitive to physics Beyond the Standard Model (BSM). 
On the other hand, sufficient theoretical control on decay amplitudes is absent in rare charm decays due to overwhelming resonance pollution and a poor convergence of the heavy quark expansion~\cite{Feldmann:2017izn}. Therefore, null test observables are inevitable for New Physics (NP) searches in $c\to u\ell^+\ell^-$ transitions, even more compelling than in the down-type counterparts $b\to s(d)\ell^+\ell^-$, where null tests already are a crucial part of ongoing and future precision programs, see Refs.~\cite{Cerri:2018ypt, Belle-II:2018jsg}.

The experimental and theoretical status of rare charm decays is presented in Ref.~\cite{Gisbert:2020vjx} and increasing and ongoing research interest ~\cite{Burdman:2001tf, Paul:2011ar, Fajfer:2012nr, Cappiello:2012vg, Wang:2014uiz, deBoer:2015boa, Fajfer:2015mia, Feldmann:2017izn, Sahoo:2017lzi, Faustov:2018dkn, DeBoer:2018pdx, Bause:2019vpr, Bause:2020obd,Bause:2020auq, Gisbert:2020vjx, Bause:2020xzj, Bharucha:2020eup, Faisel:2020php, Colangelo:2021myn, Fajfer:2021woc, Golz:2021imq} manifests the importance of progress in the field.

The plan of this talk is as follows. We present the effective field theory framework, the necessity to define null test observables in rare charm decays and give a brief overview of available upper limits in Sec.~\ref{sec:nulltest}. In Sec.~\ref{sec:baryons} we focus on the charmed baryon decay mode $\Lambda_c\to p \ell^+\ell^-$ and present angular observables, CP--asymmetries and lepton universality ratios. Sec.~\ref{sec:neutrinos} focuses on possibilities with dineutrino modes. We conclude in Sec.~\ref{sec:concl}.

\section{Effective field theory framework and null test strategies}\label{sec:nulltest}
\noindent Rare $c\to u\ell^+\ell^-$ and $c \to u \nu\bar\nu$ processes are described by the effective Hamiltonian at the charm mass scale $\mu_c$,
\begin{equation}
\mathcal{H}_{\rm eff} \supset -\frac{4G_F}{\sqrt2} \frac{\alpha_e}{4\pi} \biggl[ \sum_{k=7,9,10,S,P} \bigl( C_kO_k + C_k^\prime O_k^\prime \bigr) + \sum_{k=T,T5} C_k O_k + \sum_{ij} \bigl( C_L^{ij} Q_L^{ij} + C_R^{ij} Q_R^{ij} \bigr)\biggr]\,,
\label{eq:Heff}
\end{equation}
where the dimension 6 operators are defined as follows:
\begin{equation}
\begin{split}
O_7 &= {m_c \over e} (\ubar_L \sigma_{\mu\nu} c_R) F^{\mu\nu} \,,\quad\quad\quad O^\prime_7 = {m_c \over e} (\ubar_R \sigma_{\mu\nu} c_L) F^{\mu\nu}\,,  \\
O_9 &= (\ubar_L \gamma_\mu c_L) (\lbar \gamma^\mu \ell) \,,\quad\quad\quad\quad O^\prime_9 = (\ubar_R \gamma_\mu c_R) (\lbar \gamma^\mu \ell) \,, \\ 
O_{10} &= (\ubar_L \gamma_\mu c_L) (\lbar \gamma^\mu \gamma_5 \ell) \,, \,\,\,\,\quad\quad O^\prime_{10} = (\ubar_R \gamma_\mu c_R) (\lbar \gamma^\mu \gamma_5 \ell) \,, \\
O_S &= (\ubar_L c_R) (\lbar\ell) \,, \quad\quad\quad\quad\quad\quad O^\prime_S = (\ubar_R c_L) (\lbar\ell) \,, \\
O_P &= (\ubar_L c_R) (\lbar \gamma_5 \ell) \,, \quad\quad\quad\quad\quad O^\prime_P = (\ubar_R c_L) (\lbar \gamma_5 \ell) \,, \\
O_T &= \frac{1}{2} (\ubar \sigma_{\mu\nu} c) (\lbar \sigma^{\mu\nu} \ell) \,, \quad\quad\,\,\,\,
O_{T5} = \frac{1}{2} (\ubar \sigma_{\mu\nu} c) (\lbar \sigma^{\mu\nu} \gamma_5 \ell) \,, \\
Q_L^{ij} &=(\ubar_L \gamma_\mu c_L) (\bar \nu_{Lj}\gamma^\mu \nu_{Li})\,, \,\,\,\,\quad Q_R^{ij} =(\ubar_R \gamma_\mu c_R) (\bar \nu_{Lj}\gamma^\mu \nu_{Li})\,,
\end{split}
\label{eq:operators}
\end{equation}
with the electromagnetic field strength tensor $F^{\mu\nu}$ and $\sigma^{\mu\nu}=\frac{i}{2}\,[\gamma^\mu,\,\gamma^\nu]$. For the dineutrino operators $Q_{L\,(R)}^{ij}$ the indices $ij$ count the neutrino flavors explicitly, while we omit flavor indices for the charged leptons, as we refer to the muon case $ij=\mu\mu$, unless otherwise stated. Note that primed operators are obtained via $L\leftrightarrow R$.

\begin{figure}[h!]\centering
\includegraphics[width=0.45\textwidth]{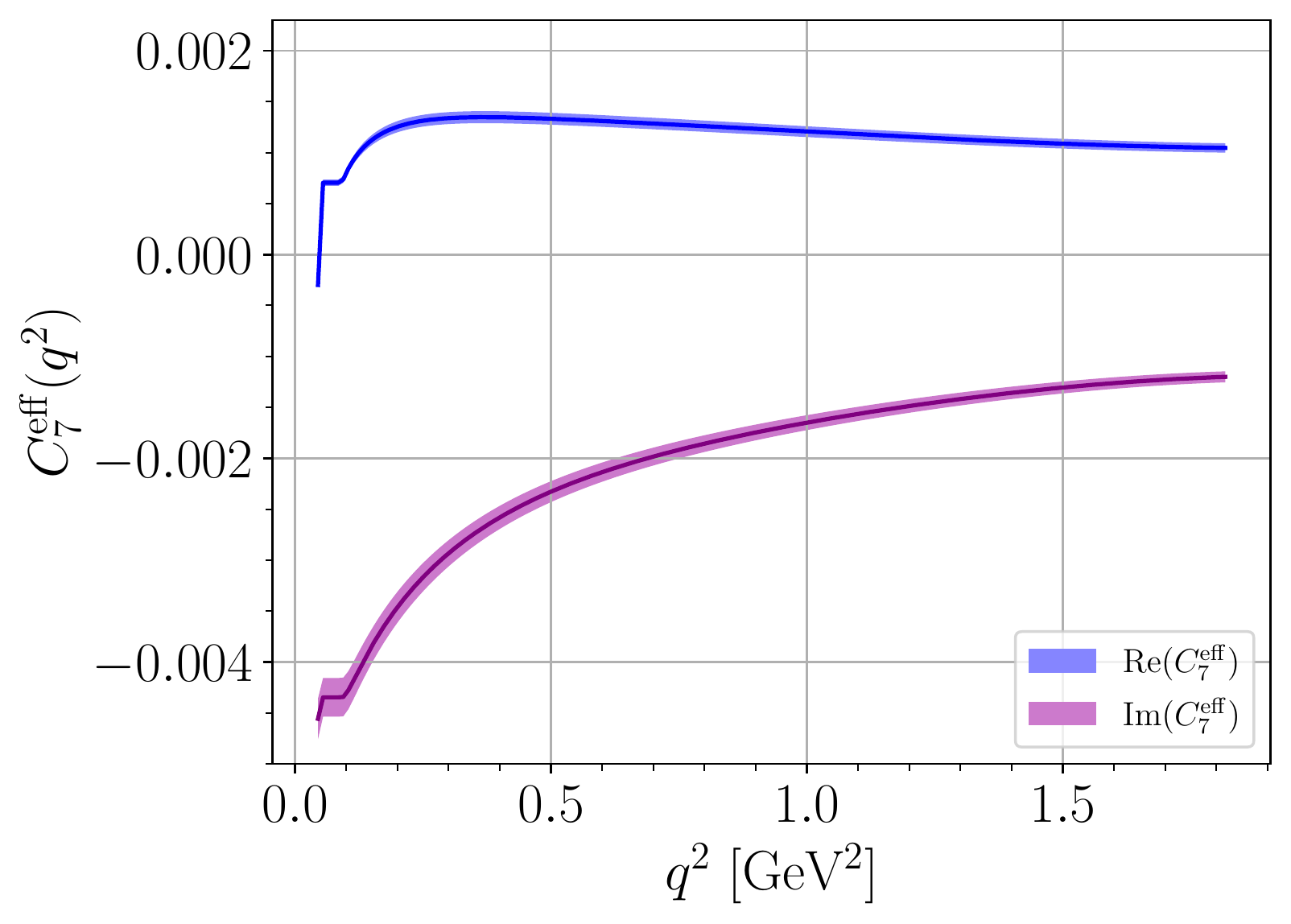}
\includegraphics[width=0.45\textwidth]{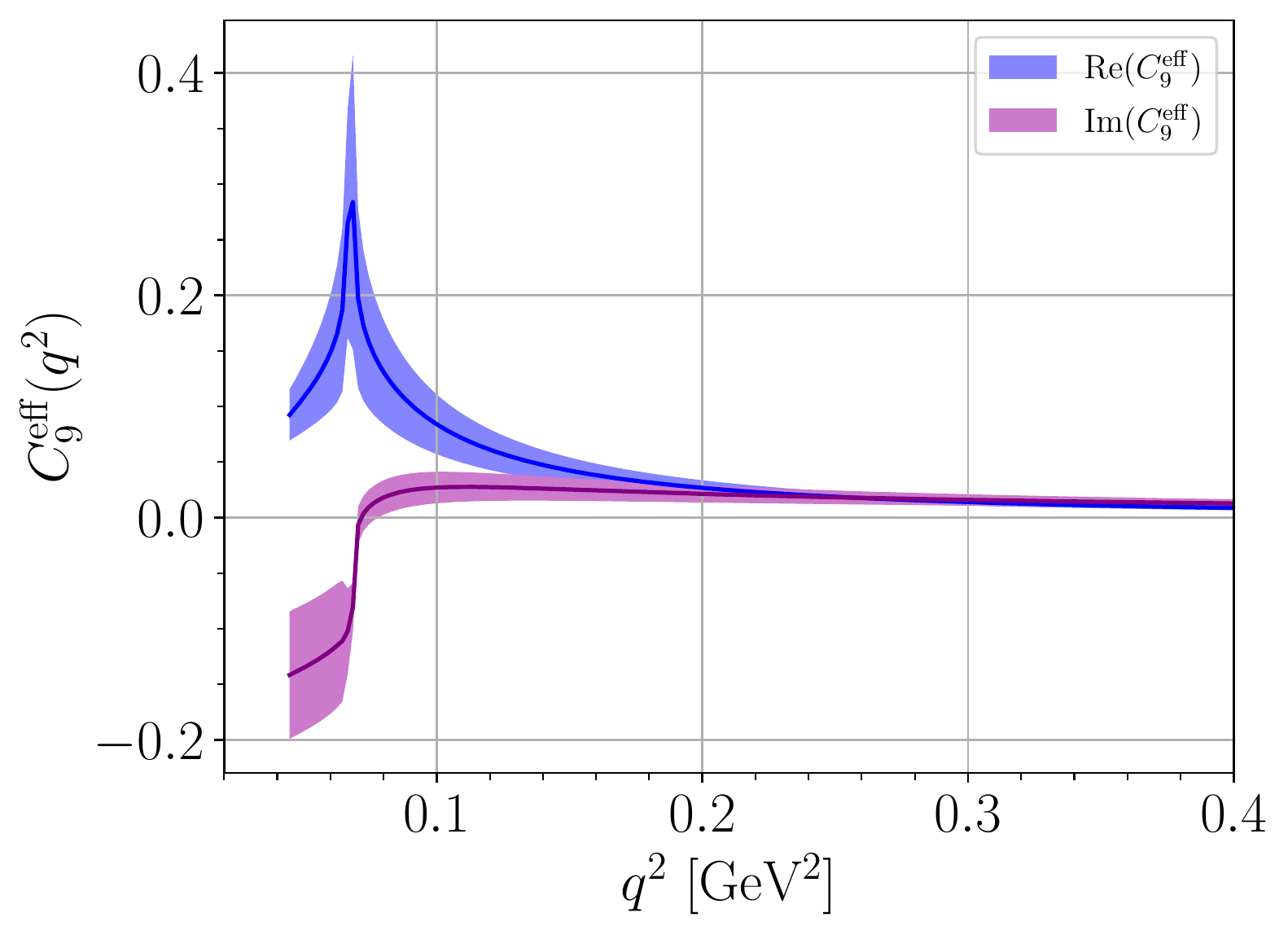}
\includegraphics[width=0.45\textwidth]{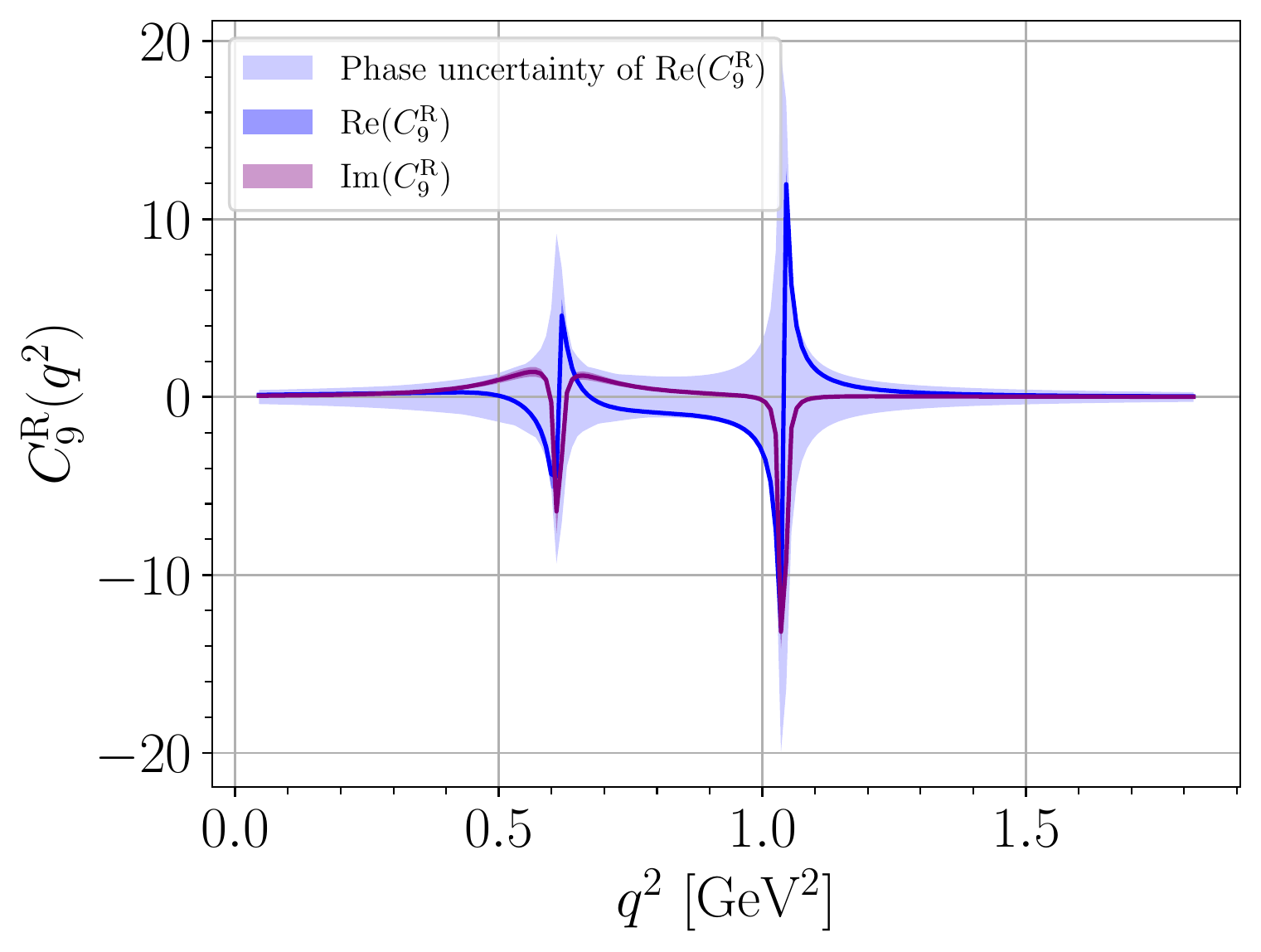}
\caption{Real and imaginary parts of the perturbative contributions $C_7^{\text{eff}}(q^2)$ (upper left plot), $C_9^{\text{eff}}(q^2)$ (upper right plot) and resonant contribution as in Eq.~\eqref{eq:resonances} (lower plot). The figure is taken from App.~A in Ref.~\cite{Golz:2021imq}, which we refer to for more details.}
\label{fig:wilsons}
\end{figure}

In order to predict branching ratios of $c\to u\ell^+\ell^-$ induced modes, one has to split the hadronic part from the short distance effects, encoded in the Wilson coefficients $C_k$ in Eq.~\eqref{eq:Heff}. Turning to predictions of SM contributions we observe that the severe GIM cancellation leads to tiny effects, reflected in small SM Wilson coefficients. In fact, all contributions, which originate solely from four quark operators at the $W$ mass scale, can be accounted for by effective $q^2$ dependent coefficients $C_7^{\text{eff}}(q^2),\,C_9^{\text{eff}}(q^2)$, where $q^2$ is the dilepton invariant mass squared. For details we refer to Refs.~\cite{deBoer:2015boa, deBoer:2017way, deBoer:thesis}. The real and imaginary parts of $C_7^{\text{eff}}(q^2),\,C_9^{\text{eff}}(q^2)$ are shown in the upper row of Fig.~\ref{fig:wilsons}.

With $C_7$ at the permille level, $C_9\sim\mathcal{O}(\%)$ for most of the kinematic region ($q^2>0.1$) and all other Wilson coefficients in Eq.~\eqref{eq:Heff} remaining zero, expected rates of rare charm decays stay beyond current experimental reach. However, $\vert \Delta c\vert=\vert\Delta u \vert = 1 $ transitions receive further contributions from intermediate resonances. These can be parametrized in terms of a sum of Breit-Wigner distributions fit to data. These resonance effects are then added as $q^2$ dependent Wilson coefficients $C_9^R(q^2),\,C_P^R(q^2)$ and are the main source of uncertainty due to unknown strong phases entering the parametrization. We exemplarily show the real and imaginary part of $C_9^R(q^2)$ for the decay mode $\Lambda_c\to p\mu^+\mu^-$ in the lower plot of Fig.~\ref{fig:wilsons} and observe a clear dominance with respect to the perturbative contributions.

The phenomenological parametrization includes the spin-1 resonances $M=\rho(770)$, $\omega(782)$, $\phi(1020)$ and relates the $\rho$ and $\omega$ contributions via an isospin relation,

\begin{equation}
  \label{eq:resonances}
  C^{R}_{9}(q^{2}) = a_{\omega}e^{\text{i}\delta_{\omega}}\left(\frac{1}{q^{2} - m^{2}_{\omega} + \text{i}m_{\omega}\Gamma_{\omega}} - \frac{3}{q^{2} - m^{2}_{\rho} + \text{i}m_{\rho}\Gamma_{\rho}}\right)
        + \frac{a_{\phi}e^{\text{i}\delta_{\phi}}}{q^{2} - m^{2}_{\phi} + \text{i}m_{\phi}\Gamma_{\phi}},
\end{equation}
where the strong phases $\delta_\omega,\,\delta_\phi$ are varied independently from $-\pi$ to $\pi$, whereas the moduli $a_\omega,\,a_\phi$ are fixed via
\begin{equation}
  \label{eq:resonance_parameters_extraction}
  \mathcal{B}_{C_9^R}(\Lambda_{c} \to  p\mu^{+}\mu^{-}) =  \mathcal{B}(\Lambda_{c} \to pM)\mathcal{B}(M \to \mu^{+}\mu^{-})\,,\quad\text{with }M=\omega,\,\phi\,.
\end{equation}
Input on branching ratios, masses and decay widths is taken from the PDG~\cite{ParticleDataGroup:2020ssz}. One finds $a_\omega=0.065\pm0.016\,$ and $a_\phi=0.110\pm0.008\,$ for the $\Lambda_c\to p \mu^+\mu^-$ case, however the same procedure can be applied to other rare charm baryon modes~\cite{Golz:2021imq}, $D\to P \ell^+\ell^-$ modes~\cite{Bause:2019vpr} and $D\to P_1P_2\ell^+\ell^-$ modes~\cite{DeBoer:2018pdx}.

\begin{figure}[h!]\centering
\includegraphics[width=0.49\textwidth]{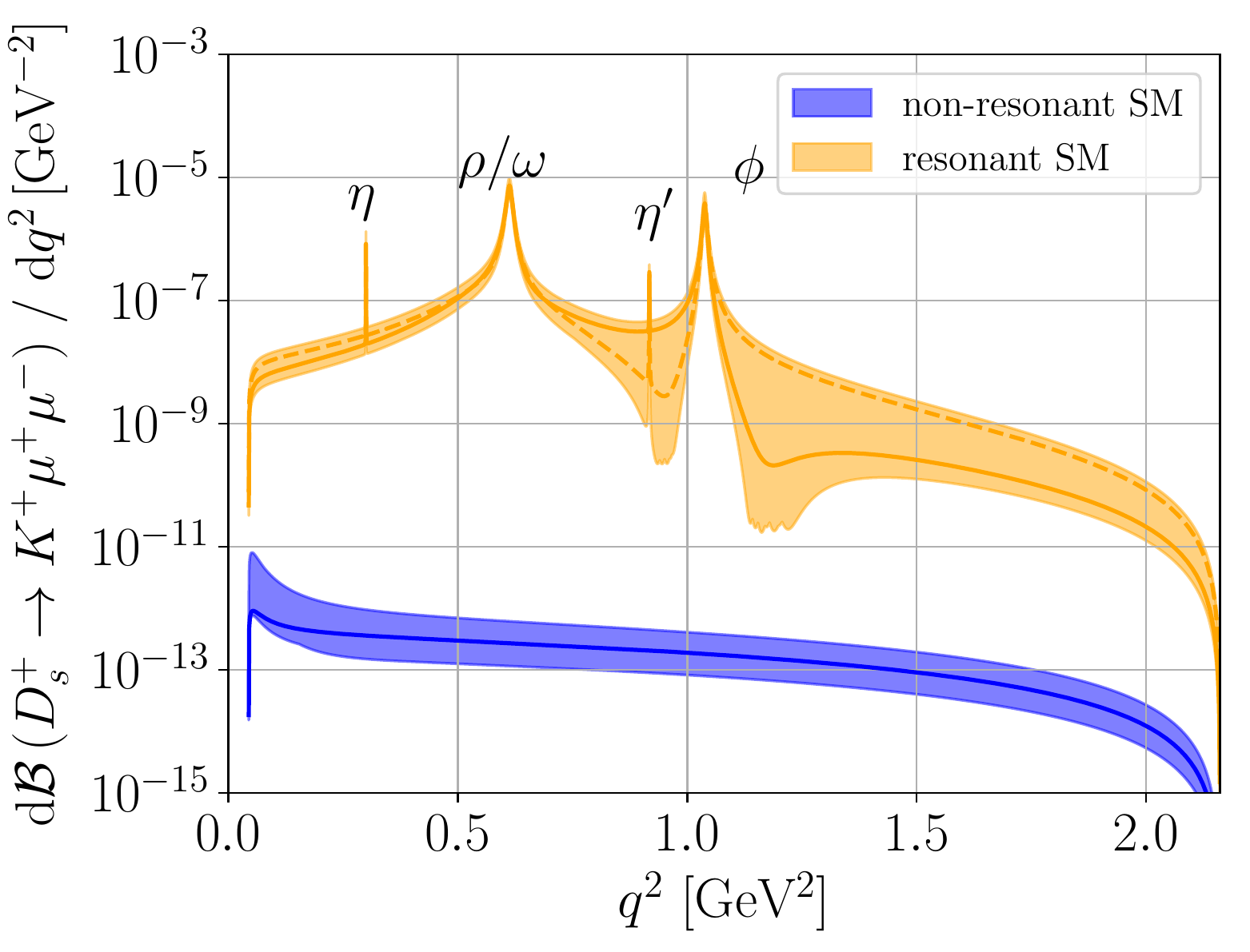}
\includegraphics[width=0.48\textwidth]{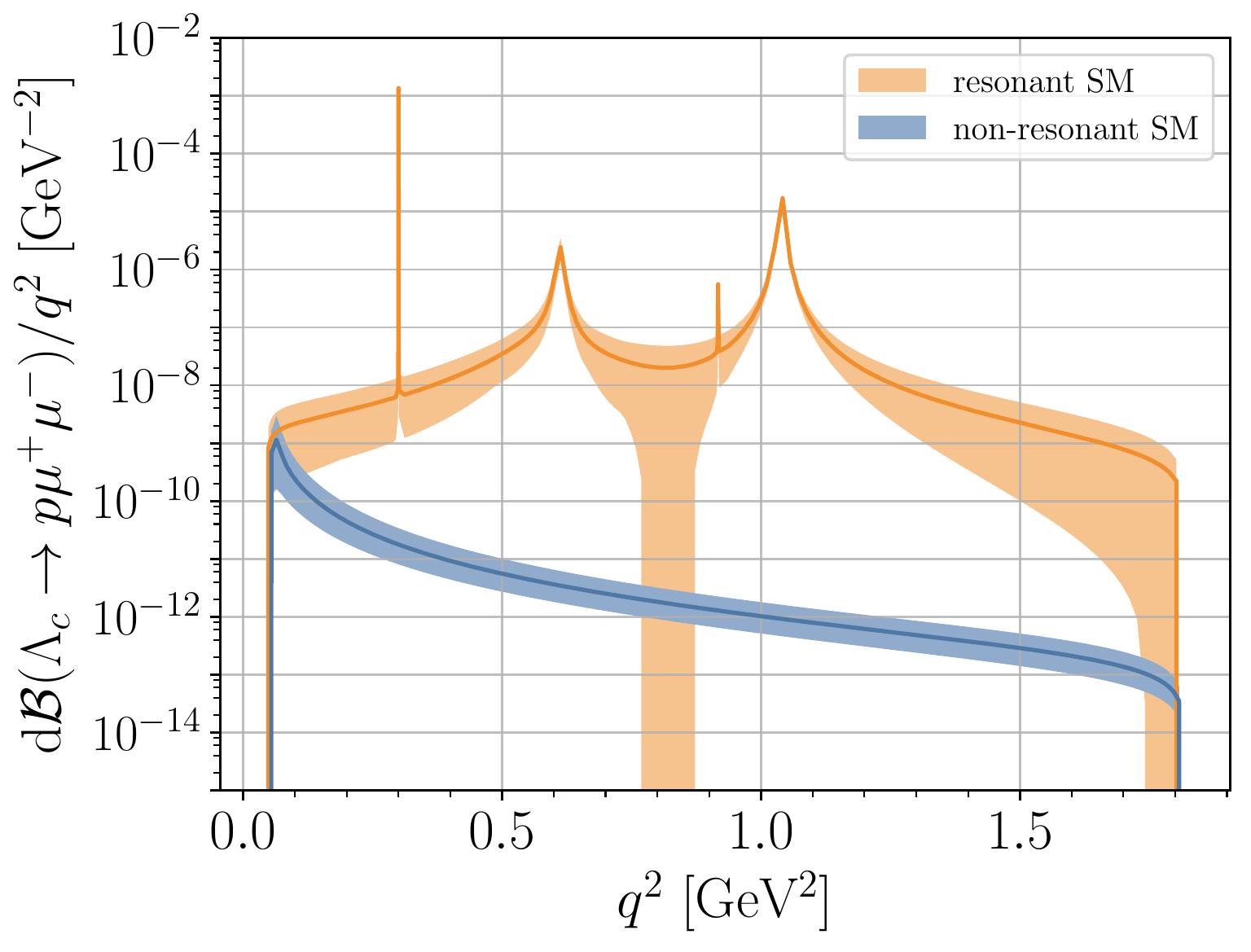}
\caption{The differential branching ratios in the SM for $D_s^+\to K^+ \mu^+\mu^-$ (left) and $\Lambda_c\to p \mu^+\mu^-$ (right). (Non-)resonant contributions are shown in orange (blue). The widths indicate uncertainties from form factors plus strong phase variation in the resonant case and charm mass scale $\mu_c$ variation in the non-resonant case. The figures are adapted from Ref.~\cite{Bause:2019vpr} (left) and Ref.~\cite{Golz:2021imq} (right).}
\label{fig:br}
\end{figure}

In Fig.~\ref{fig:br} the dominance of resonance over perturbative contributions is evident. We exemplarily show the $q^2$ differential branching ratios for $D_s^+\to K^+ \mu^+\mu^-$ (left) and $\Lambda_c \to p \mu^+\mu^-$ (right) in the SM. Here, the orange bands correspond to resonance contributions via $C_{9\,(P)}^R(q^2)$, where the main source of uncertainty is due to varying strong phases independently from $-\pi$ to $\pi$. The blue bands show the perturbative contribution obtained via $C_{7\,(9)}^{\text{eff}}(q^2)$ and varying the charm mass scale $\mu_c$. Both bands include form factor uncertainties from Lattice QCD results in Refs.~\cite{Lubicz:2017syv, Lubicz:2018rfs} for $D\to P \ell^+\ell^-$ and Ref.~\cite{Meinel:2017ggx} for $\Lambda_c\to p \ell^+\ell^-$.

Clearly, perturbative SM contributions cannot be tested in branching ratio measurements as they are shielded by resonance contributions in the full kinematic range. This, however, is exactly the motivation to study null test observables instead of branching ratios (only). Here, any signal implies the breakdown of the SM. Null tests are based on (approximate) symmetries of the SM that can be violated in BSM extensions. This includes CP-symmetry, lepton universality, lepton flavor conservation and the efficient GIM mechanism.

Experimentally, data are available for a variety of decay modes and have recently been compiled in Ref.~\cite{Gisbert:2020vjx}. Except for $D^0\to \pi^+\pi^-\mu^+\mu^-$ and $D^0\to K^+K^-\mu^+\mu^-$, only upper limits are placed for rare charm decay modes outside of the resonance search regions. However, the obtained upper limits are close to the upper ends of the resonance estimates. Additionally these upper limits imply maximally allowed values for NP contributions to Wilson coefficients in Eq.~\eqref{eq:Heff}. The bounds on dipole and muon (axial-)vector Wilson coefficients relevant in this work read
\begin{equation}
\vert C_{7}^{(\prime)}\vert \lesssim 0.3\,,\quad\vert C_{9}^{(\prime)}\vert \lesssim 0.9\,,\quad\vert C_{10}^{(\prime)}\vert \lesssim 0.8\,,
\end{equation}
which defines the model-independent playground for benchmark scenarios in null test observables studied in the remainder of this talk.

\section{Rare baryon decays}\label{sec:baryons}
\noindent In this section we discuss possibilities to test BSM physics in null test observables of ${\Lambda_c\to p\mu^+\mu^-}$ and compare our findings with other rare charm decays. We focus on angular observables in Sec.~\ref{sec:ang}, CP--asymmetries in Sec.~\ref{sec:cpv} and lepton universality ratios in Sec.~\ref{sec:lu}.

\subsection{Angular observables}\label{sec:ang}

\noindent To start off, we take a look at the fully differential angular distribution for unpolarized ${\Lambda_c\to p \mu^+\mu^-}$, conveniently written as

\begin{equation}
  \frac{\text{d}^2\Gamma}{\text{d}q^2\text{d}\cos\theta_\ell}=\frac{3}{2}\,(K_{1ss}\,\sin^2\theta_\ell\,+\,K_{1cc}\,\cos^2\theta_\ell\,+\,K_{1c}\,\cos\theta_\ell)\,,
  \label{eq:angl_distr}
\end{equation}
where $\theta_\ell$ is defined as the angle of the $\ell^+$ momentum with respect to the negative direction of flight of the $\Lambda_c$ in the dilepton rest frame. The full dependence on Wilson coefficients and form factors of the angular coefficients $K_{1ss},\,K_{1cc},\,K_{1c}$ is given in Ref.~\cite{Golz:2021imq}.

\begin{figure}[h!]\centering
\includegraphics[width=0.45\textwidth]{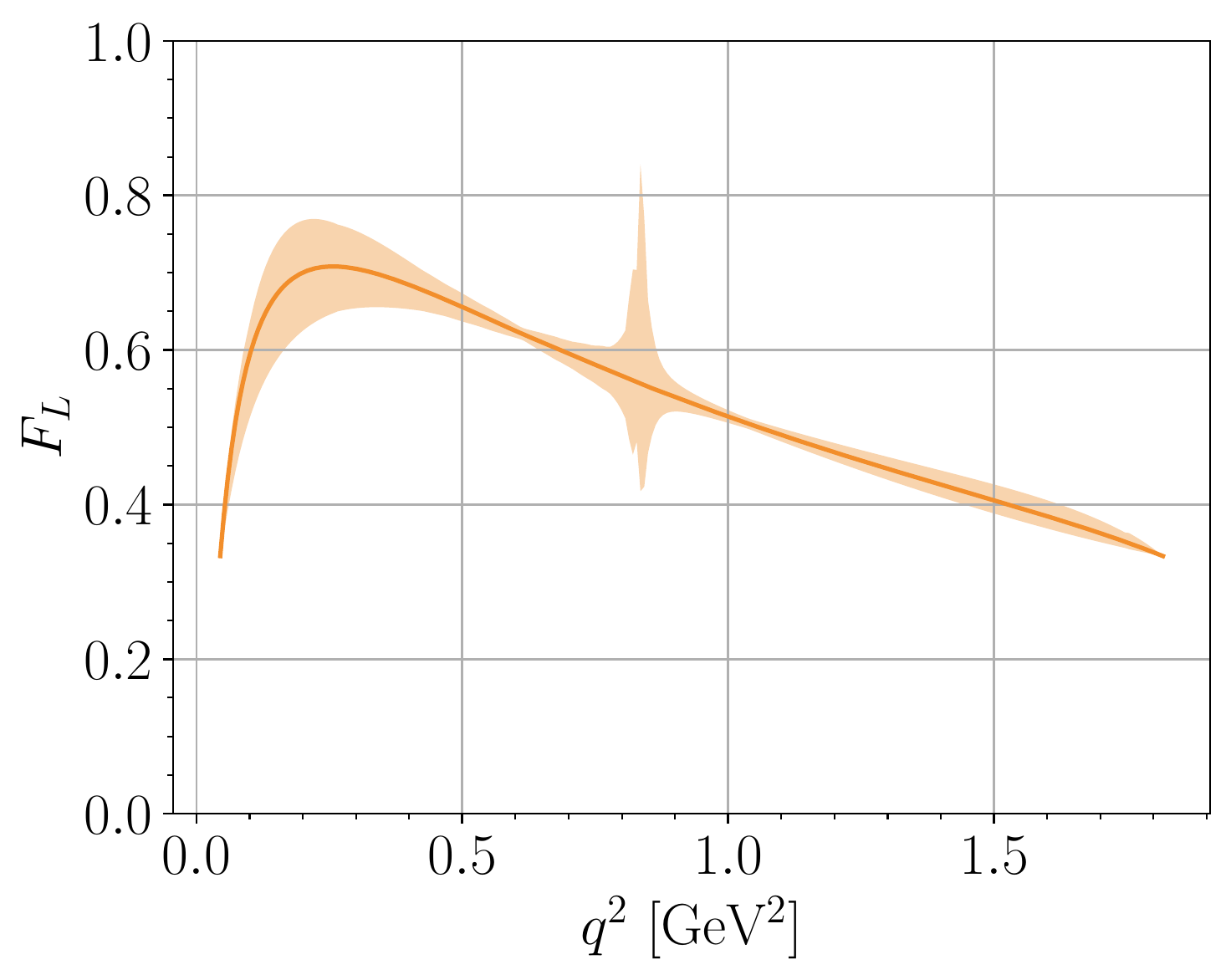}
\includegraphics[width=0.45\textwidth]{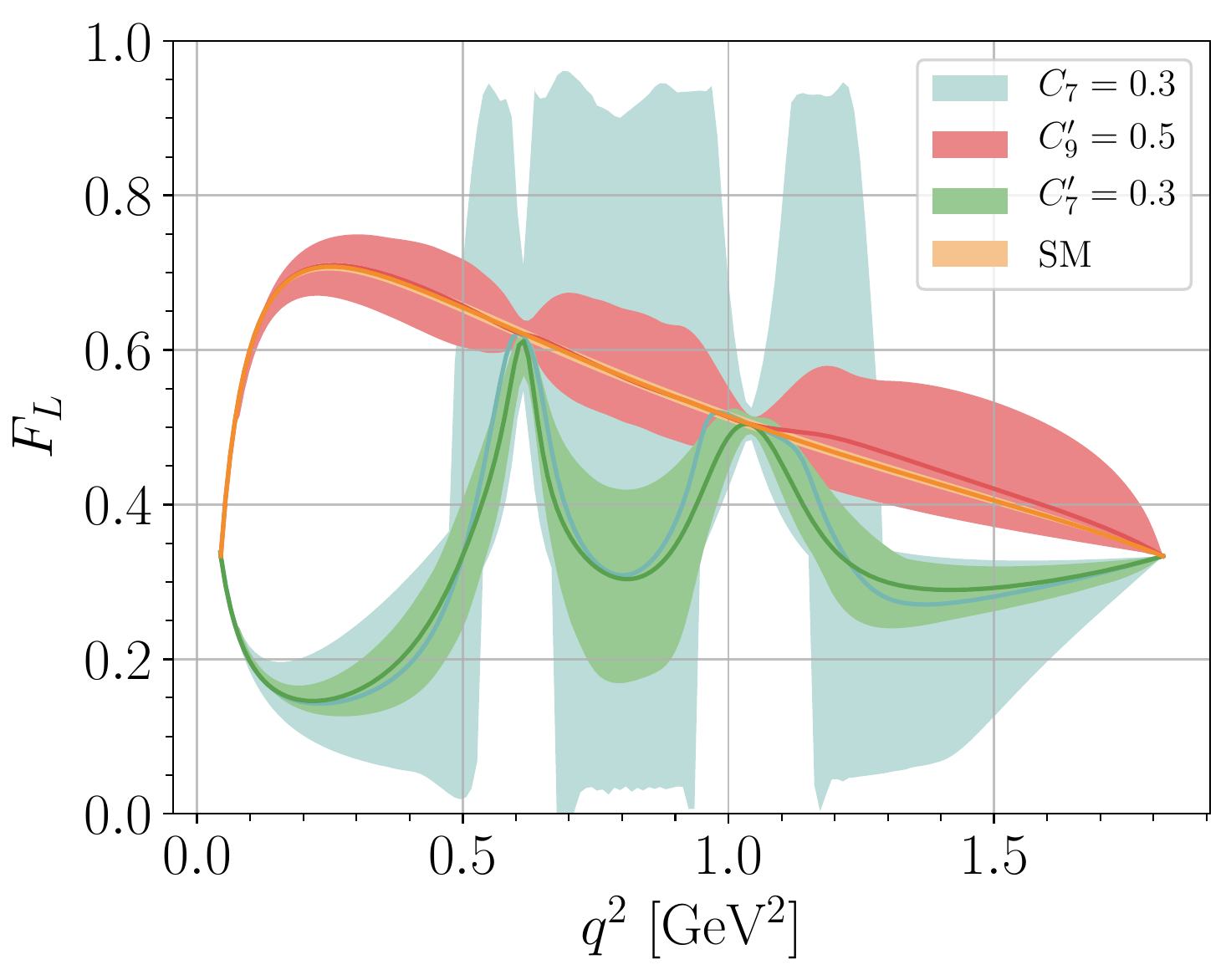}
\includegraphics[width=0.45\textwidth]{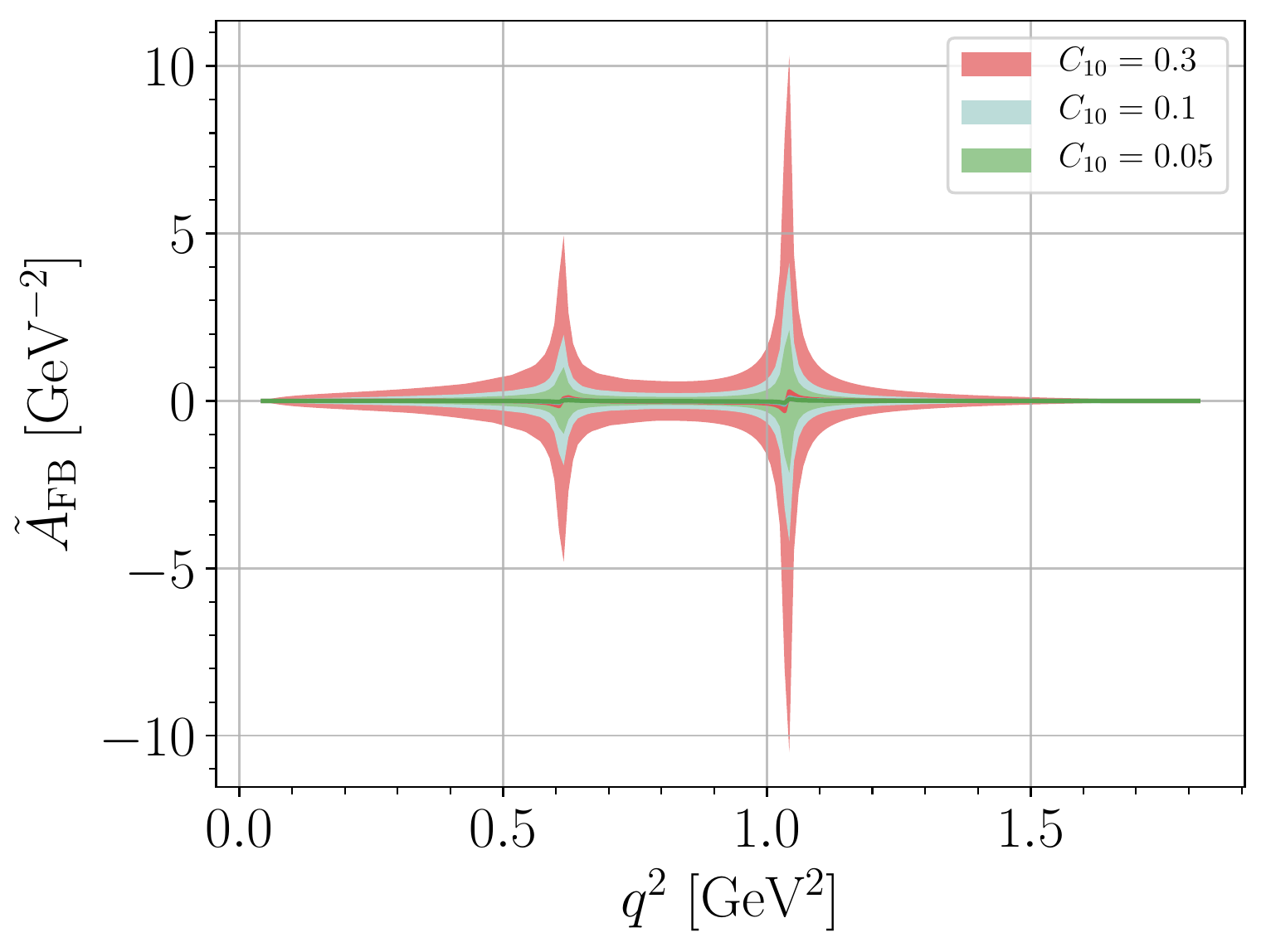}
\includegraphics[width=0.45\textwidth]{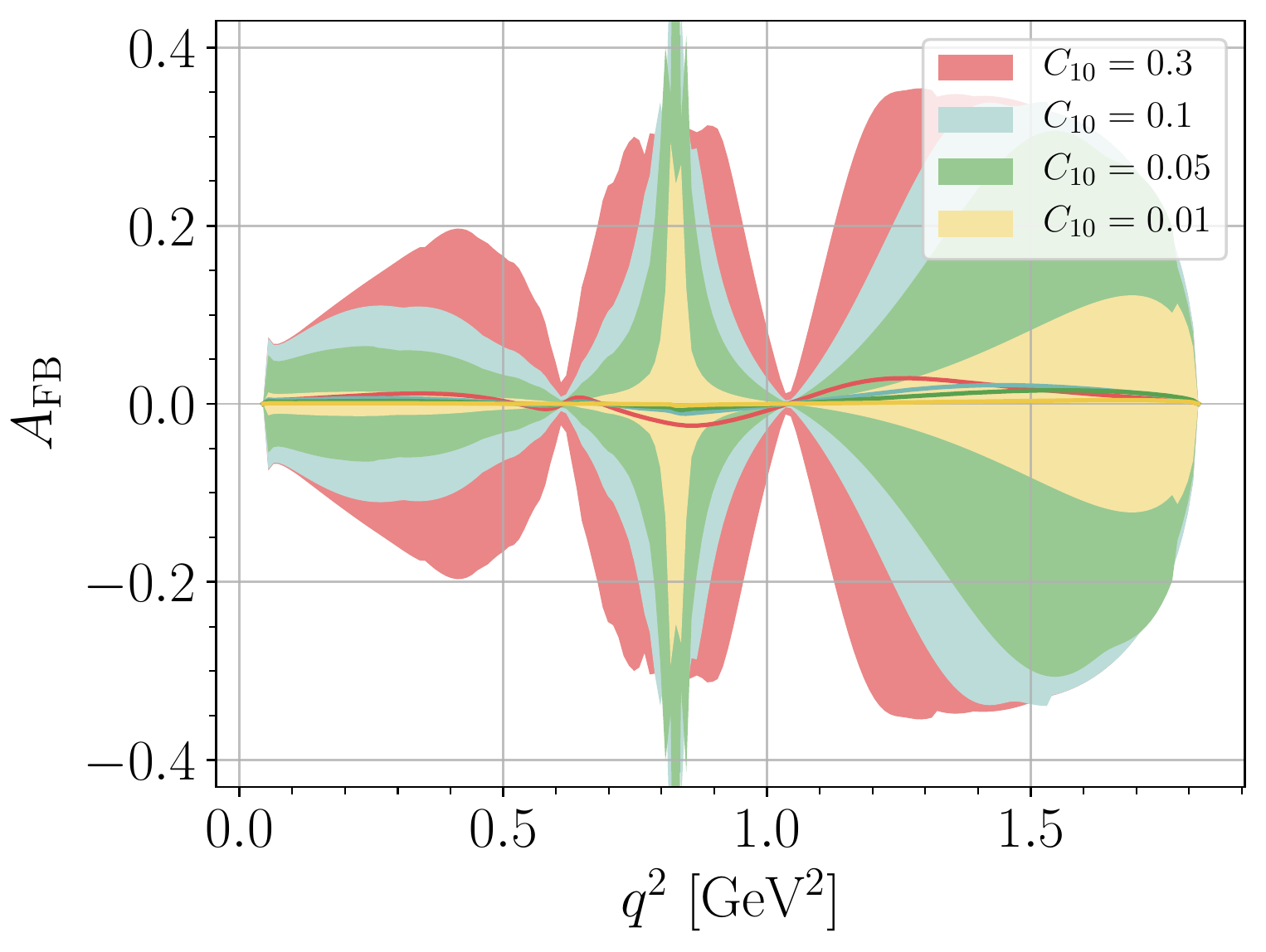}
\caption{Angular observables in $\Lambda_c\to p\mu^+\mu^-$. The upper row shows the fraction of longitudinally polarized dimuons $F_L$ in the SM (left) and in various NP scenarios (right). The lower row shows the forward-backward asymmetry $\tilde{A}_{\text{FB}}$ (left) and ${A}_{\text{FB}}$ (right) in NP scenarios with decreasing $C_{10}$ Wilson coefficient, see text for details. Plots are adapted from Ref.~\cite{Golz:2021imq}.}
\label{fig:ang}
\end{figure}

Fig.~\ref{fig:ang} includes plots of the fraction of longitudinally polarized dimuons $F_L$ and the forward-backward asymmetry $A_{\text{FB}}$ (lower row), which are the two independent observables next to the differential branching ratio that can be studied in this mode and are defined as
\begin{align}\label{eq:angulars}
F_L=\frac{2\,K_{1ss}-K_{1cc}}{2\,K_{1ss}+K_{1cc}}\,, \quad A_{\text{FB}}=\frac{3}{2}\,\frac{K_{1c}}{2\,K_{1ss}+K_{1cc}}
\end{align} 

For $F_L$ the SM expectation is shown in the left plot, where the dominant contribution stems from $C_9^R(q^2)$ and the dark orange band indicates uncertainties due to interference with $C_7^{\text{eff}}(q^2)$. The right plot shows the SM in orange and NP benchmark scenarios in blue, red and green for $C_7=0.3\,$, $C_9^\prime=0.5\,$ and $C^\prime_7=0.3\,$, respectively. Clearly, $F_L$ is not a null test per se, however, due to strong cancellations of hadronic uncertainties $F_L$ has excellent sensitivity to NP contributions. Especially NP in dipole operators leads to a different shape of the distribution, which can be understood from helicity arguments, as discussed in Ref.~\cite{Golz:2021imq}. Also see Ref.~\cite{Hiller:2021zth} for a general discussion of endpoint relations for baryons dictating $F_L(q^2_{\text{min}})=F_L(q^2_{\text{max}})=\frac{1}{3}$, exactly as in Fig.~\ref{fig:ang}.

Turning to $A_{\text{FB}}$, Fig.~\ref{fig:ang} presents NP sensitivity in the axial-vector $C_{10}$. The left plot shows $\tilde{A}_{\text{FB}}(q^2)$, where the denominator is integrated and only the numerator is differential in $q^2$. The same scenarios with both, numerator and denominator, differential as in Eq.~\eqref{eq:angulars} is shown in the right plot. In both cases the SM is not shown, because $K_{1c}$ contains interference terms proportional to either $C_9\,C_{10}$, $C^\prime_9\,C^\prime_{10}$, or $C^{(\prime)}_7\,C^{(\prime)}_{10}$. Therefore, $A_{\text{FB}}$ constitutes a clean null test with sensitivity to axial-vector couplings down to the percent level.

We have seen that the simplest observables beyond the ($q^2$ differential) branching ratios already offer possibilities to test the SM in rare charm decays despite the resonance domination. $F_L$ shows high sensitivity to dipole operators and $A_{\text{FB}}$ is the first clean null test. In addition to $A_{\text{FB}}$ in $D\to P\ell^+\ell^-$, where scalar and/or tensor NP contributions are mandatory to observe a signal, the baryon mode has signal already in the presence of $C_{10}\neq0$ alone.  Motivated by these findings we press on.

\subsection{CP--asymmetries}\label{sec:cpv}
\begin{figure}[!t]\centering
\includegraphics[width=0.43\textwidth]{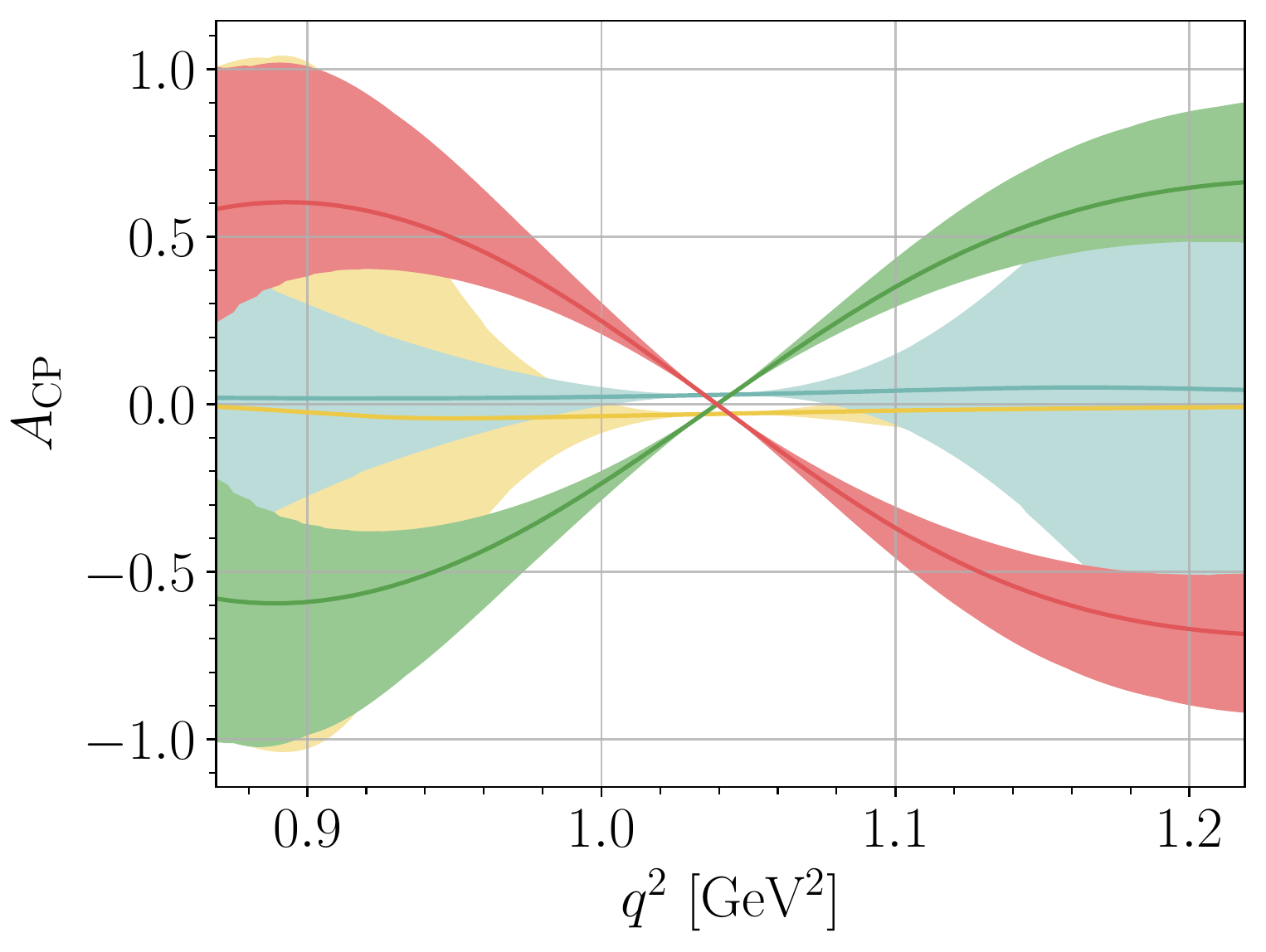}
\includegraphics[width=0.56\textwidth]{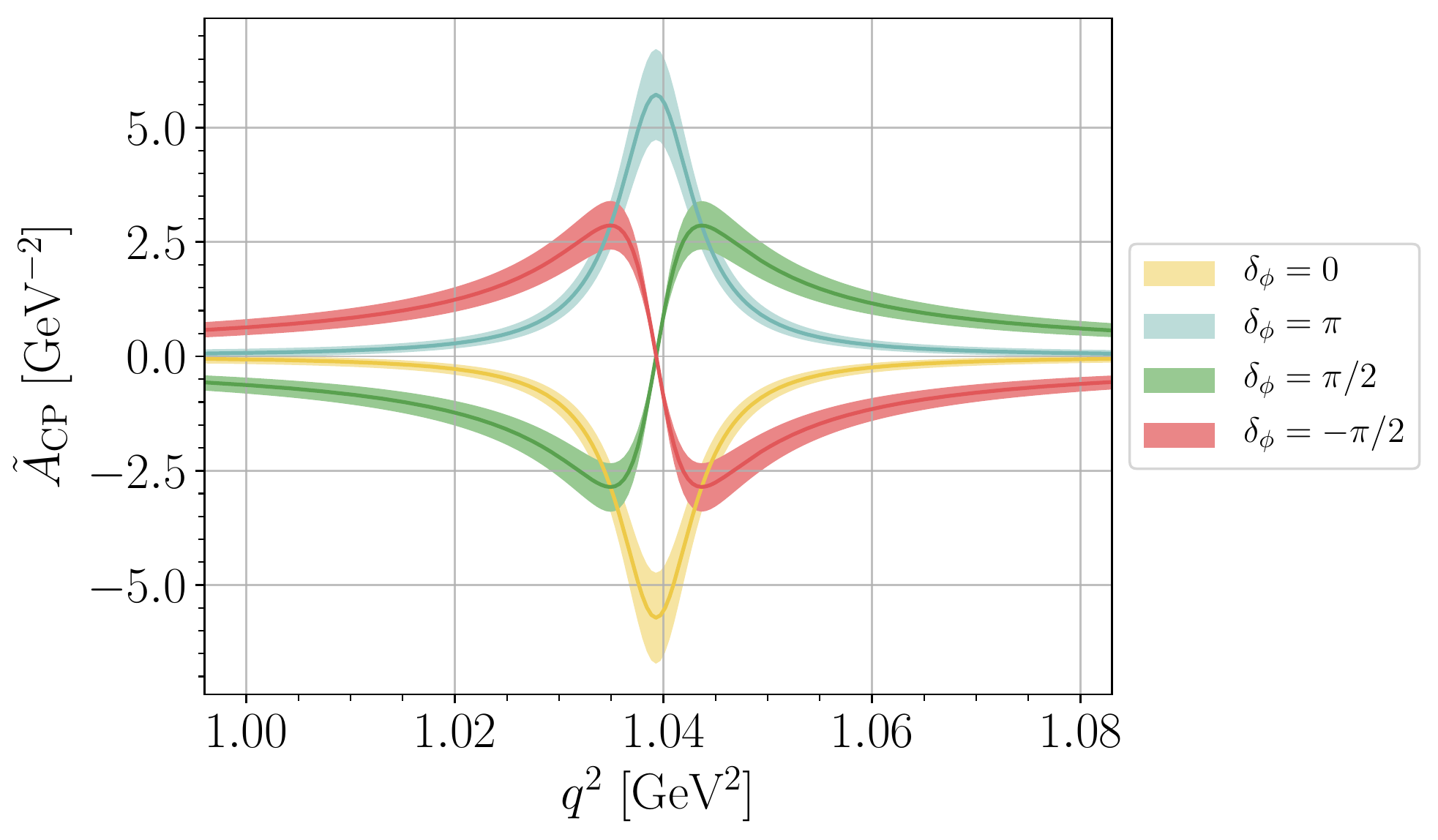}
\caption{CP--asymmetry in $\Lambda_c \to p \mu^+ \mu^-$ around the $\phi$ resonance for a $C_{9} = 0.5\text{e}^{\text{i}\pi / 4}$ benchmark for different fixed phases, see legend. The left plot displays $A_{\text{CP}}$ as in Eq.~\eqref{eq:cp}, whereas the right plot shows $\tilde{A}_{\text{CP}}$, \textit{i.e.} integrated decay rates in the denominator. Adapted from Ref.~\cite{Golz:2021imq}.}
\label{fig:cp}
\end{figure}

\noindent Following the study of angular observables, the next natural step is to investigate the CP--violating rate. 
In the left plot of Fig.~\ref{fig:cp} the CP--asymmetry
\begin{align}\label{eq:cp}
 A_{\text{CP}}=\frac{2\,K_{1ss}+K_{1cc} - 2\,\bar{K}_{1ss}-\bar{K}_{1cc}}{2\,K_{1ss}+K_{1cc} + 2\,\bar{K}_{1ss}+\bar{K}_{1cc}}\,, 
\end{align}
for  $C_{9} = 0.5\text{e}^{\text{i}\pi / 4}$ and different strong phases $\delta_\phi=0,\,\pm\frac{\pi}{2},\,\pi$ is shown in the region around the $\phi$ resonance. The barred angular coefficients $\bar{K_i}$ indicate CP--conjugation.

$A_{\text{CP}}$ decreases towards the resonance peaks, where the $\Lambda_c \to p \mu^+ \mu^-$ event rates are larger. 
This is also apparent in the right plot of Fig.~\ref{fig:cp}, which is the same as the  plot to the left but with integrated decay rates in the denominator. The CP--violating signal around the resonance increases, which is visible in $\tilde{A}_{\text{CP}}$ due to the constant denominator. Ever since the first reference in \cite{Fajfer:2012nr}, this resonance enhanced CP--violation was promoted for rare charm decays many times, see Refs.~\cite{Fajfer:2012nr, deBoer:2015boa, Fajfer:2015mia, DeBoer:2018pdx, Bause:2019vpr} as a null test of the SM, due to negligible CP--violating phases in the mixing of the first two SM quark generations. 
Next to the rate, it is also possible to study the CP-asymmetry in an angular observable. For instance, $A_{\text{FB}}^{\text{CP}}$ probes the imaginary part of $C_{10}$, complementary to the CP conserving $A_{\text{FB}}$~\cite{Golz:2021imq}.

\subsection{Lepton universality ratios}\label{sec:lu}
\noindent As a last key feature of the SM, we study lepton universality (LU) tested with $R$ ratios like
\begin{equation}
R^{\Lambda_c}_p={\int_{q^2_{\text{min}}}^{q^2_{\text{max}}} \frac{\text{d}\mathcal{B}(\Lambda_c\to p \mu^+\mu^-) }{\text{d}q^2} \text{d}q^2 }{\Big/}{\int_{q^2_{\text{min}}}^{q^2_{\text{max}}}  \frac{\text{d}\mathcal{B}(\Lambda_c\to p e^+e^-) }{\text{d}q^2} \text{d}q^2 }\,,\label{eq:R_ratio}
\end{equation}
where the same $q^2$ cuts need to be applied for both electrons and muons in the final state. LU tests in the charm sector complement similar searches in $b\to s$ transitions, where recent results already point towards a possible breakdown of the SM \cite{LHCb:2021trn}.
In $c\to u$ transitions the resonance domination leads to enhanced BSM effects, which can be revealed by exemplary BSM contributions to the muon mode in $\Lambda_c\to p \mu^+\mu^-$ leading to the results presented in Tab.~\ref{tab:R_ratios}. The SM expectation is close to one with uncertainties at the percent level. BSM effects lead to values of up to $\mathcal{O}(100)$ in the high $q^2$ regime and $\mathcal{O}(10)$ in the low $q^2$ region, whereas an integration of the full $q^2$ range yields SM-like values. The reason is again the supremacy of the resonance contributions in the full $q^2$ integral. Note that, despite the poor knowledge of these resonance effects, they result from QCD$\times$QED and therefore obey LU. Hence, similar results are obtained for LU ratios with $D\to P\ell^+\ell^-$ and $D\to P_1P_2\ell^+\ell^-$ modes~\cite{Bause:2019vpr, DeBoer:2018pdx}.

\begin{table}[h]
 \centering
  \caption{$R_p^{\Lambda_c}$~\eqref{eq:R_ratio} in the SM and in NP-scenarios with couplings to  muons for different $q^2$--bins. Ranges correspond to uncertainties and $\mathcal{O}(100)$ is indicated for the high $q^2$ region, where we only display the largest values found. Similar results are found for primed Wilson coefficients.}
 \label{tab:R_ratios}
 \begin{tabular}{l||c|c|c|c}
&SM & $\vert C^\mu_9 \vert=0.5$& $\vert C^\mu_{10}\vert=0.5$ & $\vert C^\mu_{9}\vert=\vert C^\mu_{10}\vert=0.5$ \\
 \hline
  full $q^2$ & $1.00 \pm \mathcal O(\%)$ &  SM-like & SM-like & SM-like\\
  low $q^2$ & $0.94 \pm \mathcal O(\%)$ & $7.5\ldots20$    &  $4.4\ldots13$     & $11\ldots32$   \\
  high $q^2$ & $1.00 \pm \mathcal O(\%)$ & $\mathcal{O}(100)$    & $\mathcal{O}(100)$     & $\mathcal{O}(100)$ \\
 \end{tabular}
\end{table} 

\section{Dineutrino modes}\label{sec:neutrinos}
\noindent For the last part of this talk we briefly review the possibility to test LU and charged Lepton Flavor Violation (cLFV) in the charged lepton sector utilizing experimental information on dineutrino modes. This link between the two sectors stems from the SU(2$)_L$ symmetry in the Standard Model Effective Field Theory (SMEFT) and was recently presented in Refs.~\cite{Bause:2020auq, Bause:2020xzj}.

The main feature of dineutrino branching ratios is that different contributions from Wilson coefficients of a flavor combination $\nu_i\nu_j$ need to be summed incoherently. This incoherent sum can then be rewritten as a trace over a coupling matrix in flavor space. The coupling matrices of neutrinos and charged leptons are linked via SU(2$)_L$, such that they can be replaced in the trace. Afterwards one can go back to the incoherent sum, now over charged lepton Wilson coefficients.

\begin{equation}
\begin{split}
\mathcal{B}&\propto\sum_{\nu=i,j}  \left( \vert\mathcal{C}_L^{{{U}} ij}\big\vert^2+\vert\mathcal{C}_R^{{{U}} ij}\big\vert^2 \right)=\text{Tr} \left[\mathcal{C}_L^{{{U}}}\,\mathcal{C}_L^{{{U}} \dagger}\,+\,\mathcal{C}_R^{{{U}}}\,\mathcal{C}_R^{{{U}} \dagger}\right]  \\ 
     &=  \text{Tr}  \left[ \mathcal{K}_L^{{{D}}}\mathcal{K}_L^{{{D}} \dagger}+\mathcal{K}^{{{U}}}_R\mathcal{K}_R^{{{U}} \dagger}\right]  + {\mathcal{O}}(\lambda) 
     =\sum_{\ell=i,j} \left(   \vert\mathcal{K}_L^{{{D}} ij}\big\vert^2+\vert\mathcal{K}_R^{{{U}} ij}\big\vert^2 \right)  + {\mathcal{O}}(\lambda)\,,
\end{split}
\end{equation}
with $\mathcal{B}$ being any semileptonic dineutrino branching ratio with an underlying $c\to u$ transition. Note that SU(2$)_L$ directly links the right-handed up-type Wilson coefficients, however for the left-handed operators the up-type dineutrino coefficient is linked to the down-type charged lepton coefficient and vice versa. This link is independent from the PMNS matrix and holds up to first order in Wolfenstein $\lambda$ due to the CKM rotation. However, these $\mathcal{O}(\lambda)$ corrections are taken into account in Ref.~\cite{Bause:2020xzj}.

In the following, three benchmark scenarios are put to test for the charged lepton sector via $\mathcal{K}_{L,R}^{ij}$. {Lepton-universality} (LU), {Charged lepton flavor conservation} (cLFC) and the general case with ${\mathcal{K}_{L,R}^{ij}}$ arbitrary:
\begin{align*}{
  \mathcal{K}^{\text{LU}}_{L,R}=  \left( \begin{array}{ccc}
{k} & 0 & 0 \\
0 & {k} & 0 \\
0 & 0 & {k} \end{array} \right)}\,,\quad
{  \mathcal{K}^{\text{cLFC}}_{L,R}=  \left( \begin{array}{ccc}
{k_{11}} & 0 & 0 \\
0 & {k_{22}} & 0 \\
0 & 0 & {k_{33}} \end{array} \right)}\,,\quad
{    \mathcal{K}^{\text{general}}_{L,R}=\left( \begin{array}{ccc}
{k_{11}} & {k_{12}} & {k_{13}} \\
{k_{21}} & {k_{22}} & {k_{23}}\\
{k_{31}} & {k_{32}} & {k_{33}} \end{array} \right)}\,.
\end{align*}
Constraints on charged lepton couplings are conveniently available for $c\to u\ell^+\ell^{-(\prime)}$ and $s \to d\ell^+\ell^{-(\prime)}$ couplings from searches in high-$p_T$ lepton tails of Drell-Yan $pp\to \ell^+\ell^{-(\prime)}$ transitions, see Ref.~\cite{Fuentes-Martin:2020lea, Angelescu:2020uug}. With these limits at hand, the three different assumptions on the lepton flavor structure imply different upper limits on dineutrino branching ratios. Exemplarily, we give results for two baryon modes from Refs.~\cite{Bause:2020xzj, Golz:2021imq}:
\begin{equation} \label{eq:la-nunu}
\begin{split}
\mathcal{B}(\Lambda_c^+\to p\nu\bar\nu) \lesssim 1.8\cdot10^{-6}\quad &\text{(LU)}\,,\\
\mathcal{B}(\Lambda_c^+\to p\nu\bar\nu) \lesssim 1.1\cdot10^{-5}\quad &\text{(cLFC)}\,,\\
\mathcal{B}(\Lambda_c^+\to p\nu\bar\nu) \lesssim 3.9\cdot10^{-5}\quad &\text{(general)}\,,
\end{split}
\end{equation}
and
\begin{equation} \label{eq:omeg-nunu}
\begin{split}
\mathcal{B}(\Omega_c^0\to \Xi^0\nu\bar\nu) \lesssim 3.4\cdot10^{-6}\quad &\text{(LU)}\,,\\
\mathcal{B}(\Omega_c^0\to \Xi^0\nu\bar\nu) \lesssim 1.9\cdot10^{-5}\quad &\text{(cLFC)}\,,\\
\mathcal{B}(\Omega_c^0\to \Xi^0\nu\bar\nu) \lesssim 7.1\cdot10^{-5}\quad &\text{(general)}\,.
\end{split}
\end{equation}
Measurements above the respective bounds imply the breakdown of the corresponding flavor symmetry. Since these limits are obtained in a data-driven approach, they will improve in the future. Depending on the experimental progress either the charged lepton data constrain dineutrino branching ratios or vice versa. Note that for rare charm dineutrino modes the branching ratio is a clean null test already. No resonance pollution exists in dineutrino modes and the efficient GIM cancellation yields SM expectations beyond the current and foreseeable experimental reach.

\section{Conclusions}\label{sec:concl}
\noindent We presented null test opportunities via the exemplarily chosen baryon mode $\Lambda_c\to p\mu^+\mu^-$ and commented similarities and complementarities to other rare charm decays. Dipole operators can be tested in the fraction of longitudinally polarized dimuons and axialvector couplings lead to a non-vanishing forward-backward asymmetry, with sensitivity down to the percent level in the NP Wilson coefficient $C_{10}$. While similar angular observables can be defined for the simpler meson modes $D\to P \ell^+\ell^-$, the direct sensitivity to $C_{10}$ alone highlights the benefits of baryon modes. CP--violating NP is enhanced around the resonances in all $c\to u\ell^+\ell^-$ modes, however strongly dependent on unknown strong phases. Lepton universality is tested directly in $R^{\Lambda_c}_p$ and similar ratios and indirectly via missing energy modes due to the SU(2$)_L$ link between charged leptons and neutrinos.
These null test observables overcome the problem of resonance domination in rare charm decays and provide excellent opportunities to test NP in up-type Flavor Changing Neutral Currents.

\acknowledgments
\noindent I would like to thank the organizers for the wonderful web-format of the conference. I am grateful to Rigo Bause, Hector Gisbert, Gudrun Hiller and Tom Magorsch for fruitful collaborations. A special thank goes to Rigo Bause, Gudrun Hiller and Tom Magorsch for comments on the manuscript. This work is supported by the \textit{Studienstiftung des Deutschen Volkes}.

\end{document}